\documentstyle[11pt,newpasp,twoside,psfig]{article}
\begin{document}
\title{On the pre-RLO spin-orbit couplings in LMXBs}
\author{Thomas M. Tauris}
\affil{Nordic Institute for Theoretical Physics (NORDITA),
       Blegdamsvej 17, DK-2100 Copenhagen {\O}, Denmark}

\begin{abstract}
We investigate the effect of orbital decay caused by nuclear expansion 
of a (sub)giant star in synchronous binary system. We compare
this effect with the presence of a magnetic stellar wind and show
that the additional transfer of orbital angular momentum into spin
angular momentum is relatively important -- especially since it
has been shown that the effect of magnetic braking
saturates at short orbital periods.
\end{abstract}

\section{Introduction}

The presence of magnetic stellar winds has long been known to decelerate
the rotation of low-mass stars (e.g. Kraft 1967; Skumanich 1972; 
Sonderblom 1983). The magnetized wind causes loss of spin angular 
momentum as a result of chromospheric coronal activity of cool
$< 1.5\,M_{\odot}$ stars with subphotospheric convection zones 
(Mestel 1984).
In tight synchronized binaries, the loss of spin angular momentum 
occurs at the expense of the orbital angular momentum. As a result
the orbital period decreases while the stellar components spin up
and approach one another.
Based on Skumanich's observations, Verbunt \& Zwaan (1981) derived 
an expression for the effect of magnetic braking and applied it to 
LMXBs by extrapolating, the dependence of the magnetic braking
on the orbital angular velocity, down to very short orbital periods
($\sim 1$ hr -- 1 day).
However, a fundamental law of angular momentum loss is unknown for rapidly 
rotating stars. Rappaport, Verbunt \& Joss~(1983)
investigated a weaker dependency on the stellar radius.
Meanwhile, it now seems that the necessary stellar activity perhaps
saturates for rotation periods shorter than 2--3 days
(Rucinski 1983; Vilhu \& Walter 1987; Stepien 1991)
which leads to a much flatter dependence of the angular momentum 
loss rate on the angular velocity than is given by the Skumanich-law
($\Omega ^{1.2}$ vs. $\Omega ^3$).
Based partly on observational work, Stepien~(1995) derived a new 
magnetic braking law
which smoothly matches the Skumanich-law dependence 
for wide systems to the dependence obtained by
Rucinski~(1983) for the short orbital period ($\le 3$ days) systems.

Given the new and possibly less dominant role of a magnetic stellar wind,
it is important to investigate the dependence of orbital decay on the
internal structural changes of a star in a close binary system. 
The expansion of a (sub)giant star will also cause it to rotate slower
and hence, for a synchronous system, additional transfer of orbital
angular momentum into spin angular momentum must take place. 
Here we investigate the effect of orbital decay caused by nuclear expansion 
of a pre-RLO star and compare it to various magnetic braking laws.

\section{Results}

In Figs.~1--2 we show the pre-RLO evolution of a binary consisting of a
$1\,M_{\odot}$ star and a $1.3\,M_{\odot}$ neutron star with an initial
orbital period of 3.0 days (i.e. a typical progenitor system for an LMXB).
In this case we have not included the effect of magnetic braking (which will
be included later) in order to concentrate on the spin-orbit effect
only caused by stellar expansion. Nor have we included the direct loss 
of orbital angular momentum as a result of gravitational wave radiation
and stellar wind mass loss -- both which are insignificant in this case
and omitted for the sake of clarity. The neutron star can be considered
as a point mass which only interacts with its companion through gravitation.
In each panel we have plotted a relevant parameter as a function of the
model number in our evolutionary sequence. We refer to 
Tauris \& Savonije~(1999) for further details on the stellar evolution
code and the estimate of the tidal torque and dissipation of energy.
The gray area on the right-hand side in each panel marks the onset of
the epoch of Roche-Lobe overflow (RLO). We have only
included the first hundred models or so of the mass transfer phase.\\
In Table~1 is written the values of most of the parameters for model
numbers: 1, 105 and 988 -- corresponding to ZAMS, $t=10.39$ Gyr (complete
termination of core {H}-burning) and $t=12.29$ Gyr (onset of RLO),
respectively.

\begin{figure}
\psfig{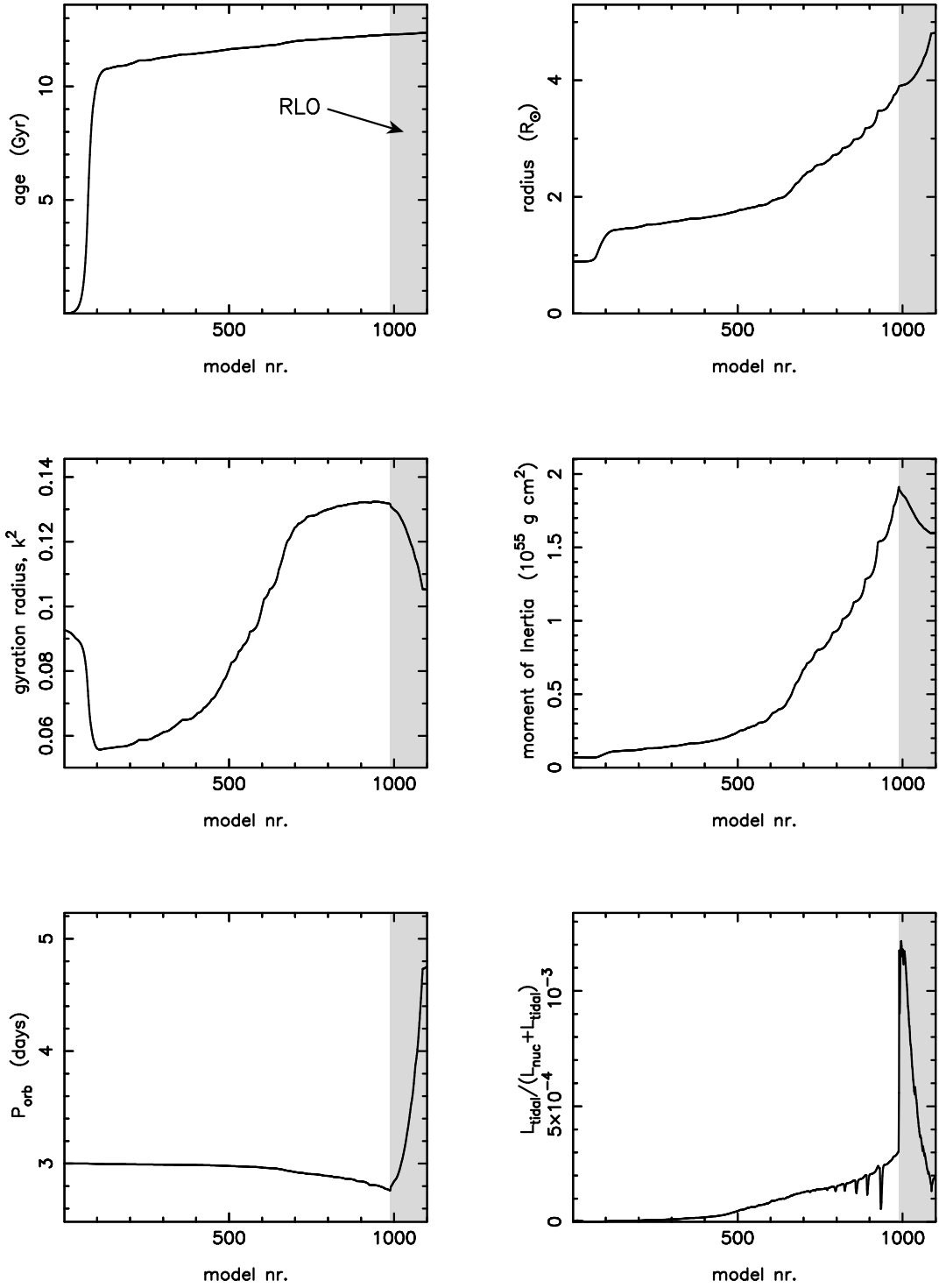}
\caption{Stellar and orbital parameters prior to RLO for a system
with a $1\,M_{\odot}$ star + $1.3\,M_{\odot}$ neutron star
and $P_{\rm orb}=3.0$ days.}
\end{figure}

\begin{figure}
\psfig{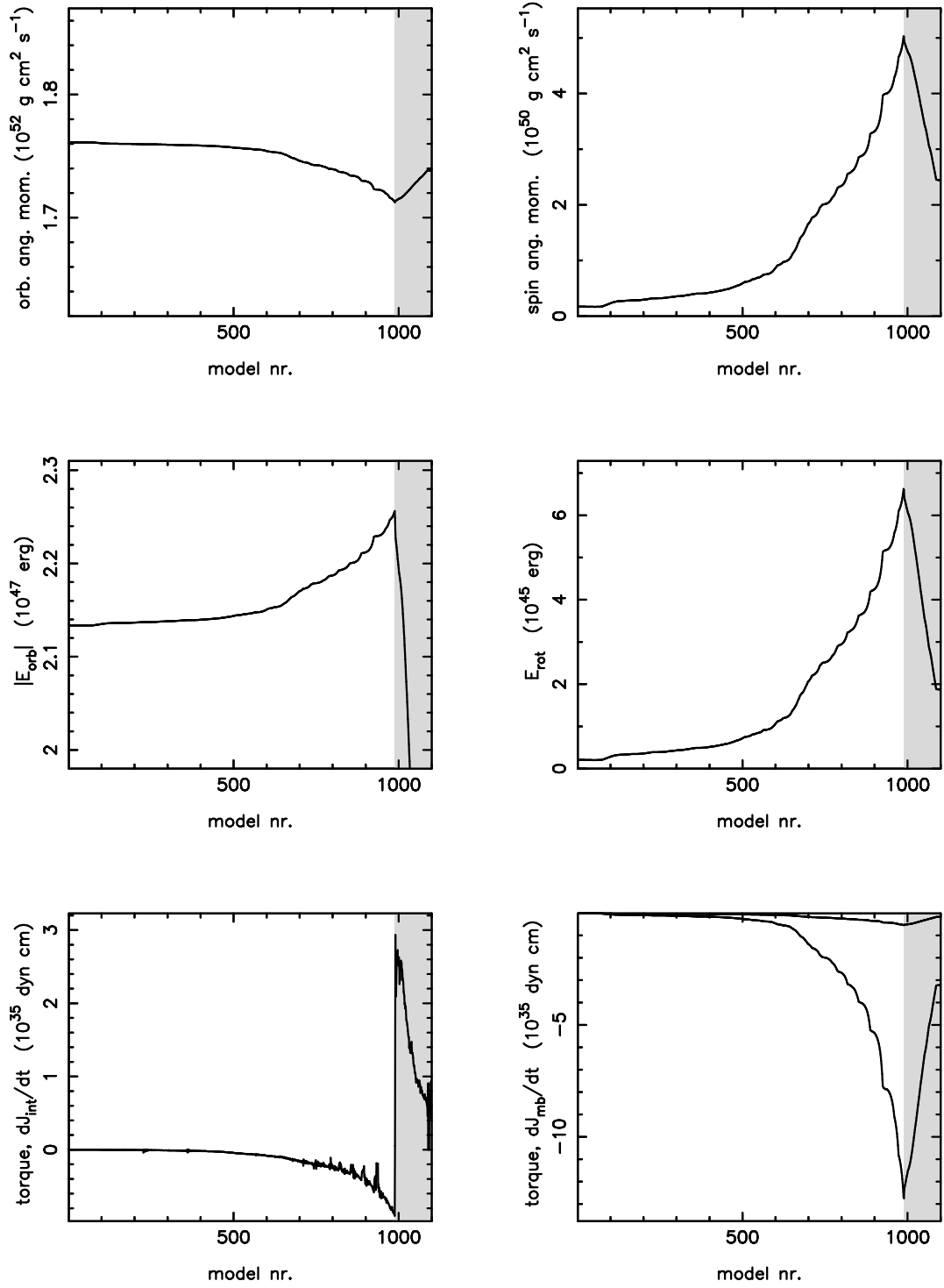}
\caption{Same system as shown in Fig.~1.}
\end{figure}

   \begin{table}
      \caption[]{Stellar and orbital parameters for the LMXB in Figs.~1--2}
         \begin{center}
         \begin{tabular}{lrrr}
           \smallskip
           \smallskip
                           & model 1 & model 105 & model 988\\
           \smallskip
           age, $t$ (Gyr)              & 0.0 & 10.39 & 12.29\\
           \smallskip
           orbital period, $P_{\rm orb}$ (days)  & 3.0000 & 2.996 & 2.759\\ 
           \smallskip
           radius, $R$ ($R_{\odot}$)   & 0.8898 & 1.364 & 3.877\\
           \smallskip
           gyration radius, $k^2$ & 0.092555 & 0.055796 & 0.13168\\
           \smallskip
           moment of Inertia, $I$ ($10^{55}$ g cm$^2$) & 0.070608 
                              & 0.10003 &1.9068\\
           \smallskip
           angular velocity, $\Omega$ ($10^{-5}$ rad s$^{-1}$) & 2.42407
                   &  2.42706 & 2.63610\\
           \smallskip
           spin ang. mom., $J_{\rm spin}$ ($10^{50}$ g cm$^2$ s$^{-1}$)
                            & 0.1712 & 0.2428 & 5.027\\ 
           \smallskip
           orbital ang. mom., $J_{\rm orb}$ ($10^{52}$ g cm$^2$ s$^{-1}$)
                            & 1.7611 & 1.7604 & 1.7125\\ 
           \smallskip
           rotational energy, $E_{\rm rot}$ ($10^{45}$ erg) & 0.20745 &
                               0.29461 & 6.6252\\
           \smallskip
           orbital energy, $E_{\rm orb}$ ($10^{47}$ erg) & 2.13359 &
                               2.13533 & 2.25627\\
           \smallskip
           orbital separation, $a$ ($R_{\odot}$) & 11.5498 &
                               11.5404 & 10.9218\\
           \smallskip
           "internal" torque, $dJ_{\rm int}/dt$ ($10^{35}$ dyn cm) & 
                                   ------ & $-$0.00168 & $-$0.9032\\
           \smallskip
           magnetic torque$^{*}$, $dJ_{\rm mb}/dt$ ($10^{35}$ dyn cm) & 
                                  $-$0.01928 & $-$0.06445 & $-$11.740\\
           \smallskip
           magnetic torque$^{**}$, $dJ_{\rm mb}/dt$ ($10^{35}$ dyn cm) & 
                                   $-$0.01748 & $-$0.02482 & $-$0.5071\\
           
         \end{tabular}
         \end{center}
          \begin{list}{}{}
         \item[$^{*}$]  Skumanich-law 
         \item[$^{**}$] Stepien (1995) 
         \end{list}
   \end{table}

We shall now briefly describe the evolution of the system considered.
While on the main-sequence (the first 10 Gyr) the gyration radius is
decreased by 40$\,$\% which causes the moment of inertia to decrease
slightly since the radius of the star is almost constant. Only toward
the very end of core {H}-burning the radius expands enough to increase
the moment of inertia. From here on, between model 105 and the onset
of RLO at model 988, the gyration radius is increased by a factor of
2.36 while the nuclear expansion of the star increases its radius by
a factor of 2.84. Hence the moment of inertia ($I=k^2\,MR^2$)
is increased by a factor of 19.0. This causes the spin angular 
momentum of the star ($J_{\rm spin}=I\Omega$) to increase similarly.
However, since the system is tidally locked the forced synchronization
compensates for the increase in spin angular momentum by tapping
orbital angular momentum and thus decreasing the orbital period (energy)
and increasing the orbital angular velocity, $\Omega$ (equal to
the stellar angular spin rate). Therefore $J_{\rm spin}$
increases by a factor of 20.7 (and not 19.0) since $\Omega$ is increased 
slightly by 8.6$\,$\%. The increase in $J_{\rm spin}$
($4.78 \times 10^{50}$ g cm$^2$ s$^{-1}$) is exactly equal to the
decrease in $J_{\rm orb}$ and hence the total angular momentum is
conserved, as it should be.
The orbital decay is seen in the decrease of
the orbital period from 3.0 days to 2.759 days and the orbital
separation is shortened by $0.6\,R_{\odot}$.
The corresponding decrease in orbital energy (increase in binding energy)
is $1.21 \times 10^{46}$~erg. However, the increase in rotational
energy ($E_{\rm rot} =\frac{1}{2}\,I\Omega^2$) of the star is only 
$6.33 \times 10^{45}$~erg and hence we can
conclude that roughly half of the liberated orbital energy is converted
into rotational energy of the non-degenerate star and roughly half of
the energy goes into heat (tidal dissipation energy) in its convective
envelope. Note, that the tidal dissipation energy rate reaches a
value of $\sim 0.1\,$\% of the total nuclear energy production rate in the
{H}-shell of the star at the moment of RLO 
($L_{\rm tidal} \sim 10^{-3}\,L_{\rm nuc}$). This is shown in the plot
in the bottom right panel of Fig.~1.

\subsection{Dependence on initial $P_{\rm orb}$ and mass}
\begin{figure}
\psfig{file=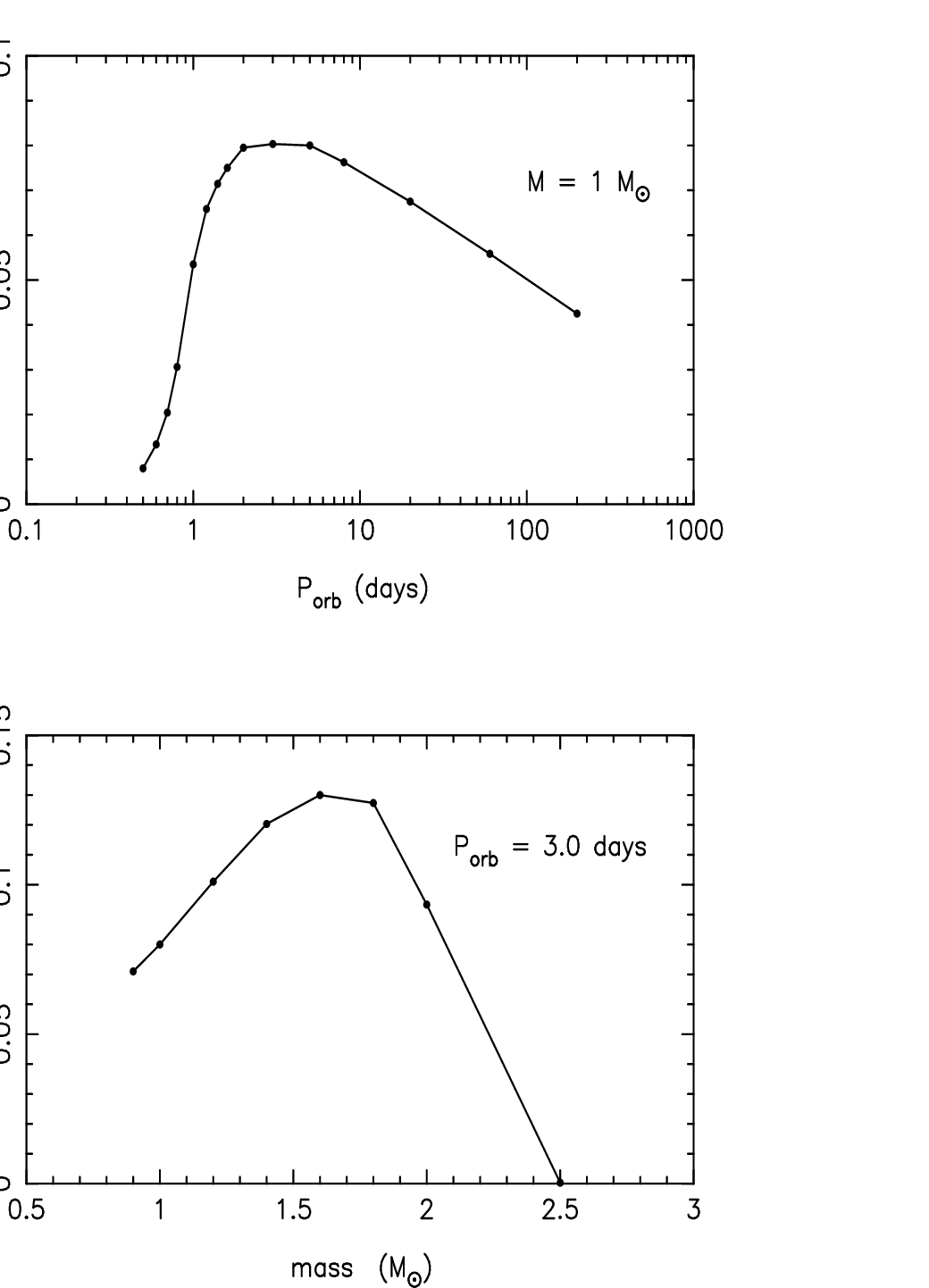,height=20.0cm,width=17.0cm}
\caption{The pre-RLO orbital decay as a function of $P_{\rm orb}$ and mass}
\end{figure}
In Fig.~3 we have plotted the relative decrease in $P_{\rm orb}$ as a function
of initial $P_{\rm orb}$ (top panel) and as a function of stellar mass
(bottom panel). The orbital period decay caused by nuclear expansion
is seen to be strongest for systems with $2\le P_{\rm orb} \le 5$ days 
(for a $1\,M_{\odot}$ star). This trend is expected, since for
very close binaries ($P_{\rm orb} < 1$ day) the star
doesn't expand much before filling its Roche-lobe. And for very wide
systems the decay also decreases since the tidal torques become weaker with
distance. Note, that for wide system ($P_{\rm orb} > 100$ days) the direct
stellar wind mass loss (not included here) would cause the systems to {\em widen}
prior to RLO since the donor star will be quite evolved before filling its
Roche-lobe (see Fig.~6 in Tauris \& Savonije 1999).\\
The orbital decay for binaries with $P_{\rm orb}=3.0$ days
is seen to be largest ($\sim$ 13$\,$\%) for stars with masses
$1.5 < M/M_{\odot} < 1.8$. This is interesting since the effect of a magnetic
stellar wind is somewhat weak for these relatively heavy stars which don't 
have very dominant convective envelopes like the $1\,M_{\odot}$ stars.
Hence, for stars in this mass interval the effect of orbital decay
due to nuclear expansion is particularly important.

\section{Magnetic braking torque}
We have demonstrated how the nuclear expansion of a binary star will cause 
the orbital period to decrease by $\sim 10\,$\% prior to RLO. In this section
we will now compare that result with the effect of magnetic braking.

The Skumanich-law expression can be written as (Verbunt \& Zwaan 1981):
\begin{equation}
  \frac{dJ_{\rm mb}}{dt} \simeq -0.5\times 10^{-28}\; f^{-2}\,k^2MR^4\,\Omega^3
           \quad \mbox{dyn cm}
\end{equation}
Stepien (1995) derived a magnetic braking law which contains the Skumanich-law
dependence ($\dot{J}_{\rm mb} \propto \Omega ^3$) for wide systems,
and with the dependence obtained e.g. by Rucinski~(1983) for short orbital systems
($\dot{J}_{\rm mb} \propto \Omega ^{1.2}$). We slightly rewrite the expression by 
Stepien (1995) and obtain (in cgs units):
\begin{equation}
  \frac{dJ_{\rm mb}}{dt} \simeq -1.90\times 10^{-16}\; k^2MR^2\,\Omega\,
            e^{-1.50\times 10^{-5}/\Omega} \quad \mbox{dyn cm}
\end{equation}
The two equations above represent a strong and a weak magnetic braking torque, respectively.
Rappaport, Verbunt \& Joss (1983) introduced a braking law depending explicitly on the
radius of the star: 
\begin{equation}
  \frac{dJ_{\rm mb}}{dt} \simeq -0.5\times 10^{-28}\; f^{-2}\,k^2MR^4_{\odot}
           \left( \frac{R}{R_{\odot}} \right) ^\gamma\,\Omega^3
           \quad \mbox{dyn cm}
\end{equation}

\begin{figure}
\psfig{file=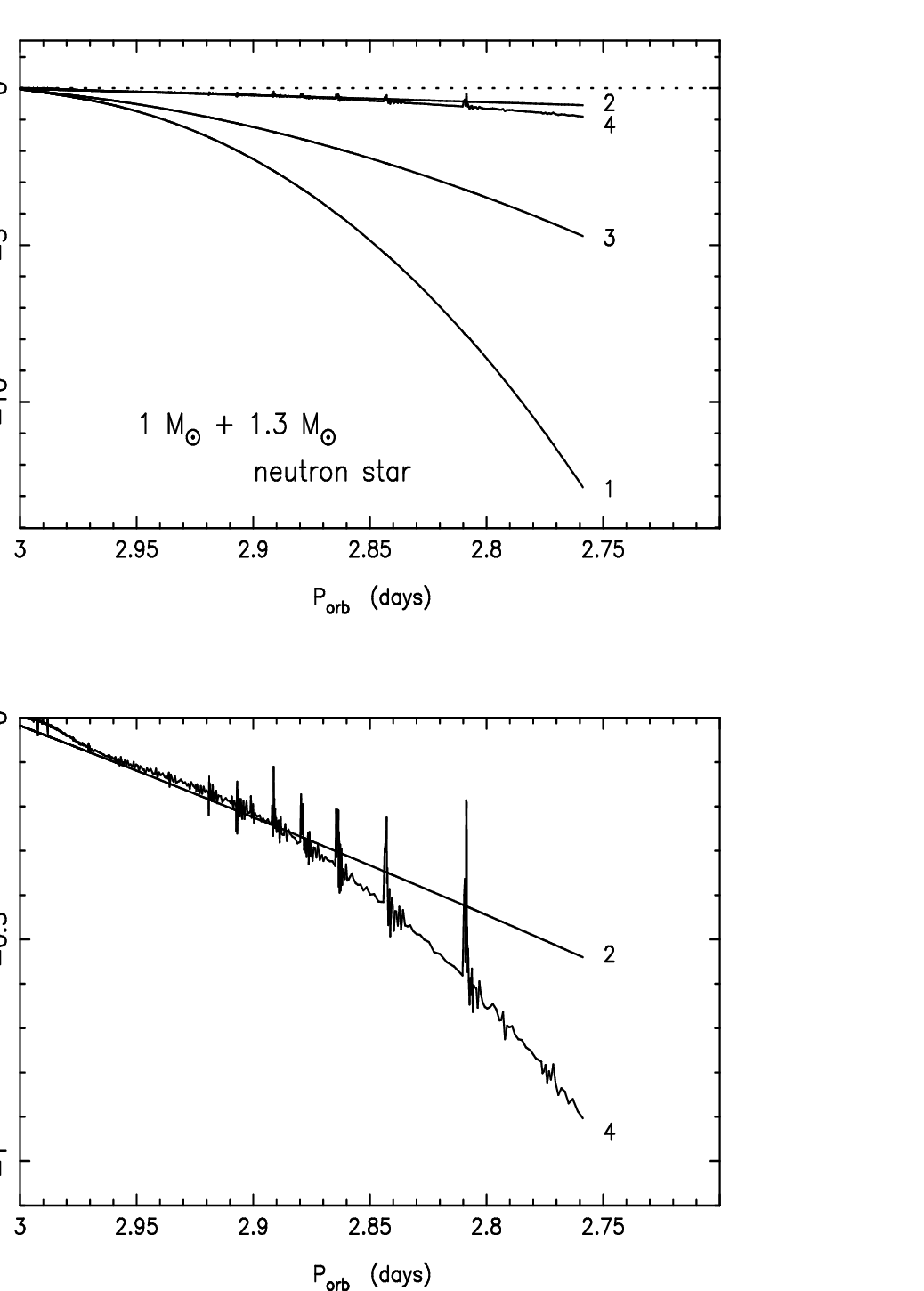,height=20.0cm,width=17.0cm}
\caption{Braking torques acting on the system shown in Figs.~1-2.}
\end{figure}

In Fig.~4 we have shown the magnitude of the different braking laws for the parameters
of the system plotted in Figs.~1-2 (i.e. for a given orbital period we have used the
values of $R$ and $k^2$ calculated in our model to estimate the torques at that exact
epoch of evolution). We assumed $\gamma = 3$ in the equation by Rappaport, Verbunt \& Joss~(1983)
for an intermediate magnetic braking law.
The label at the end of each plotted curve in the figure corresponds to the above eqs.~1--3:
(1) Skumanich-law, (2) Stepiens law, (3) RVJ.
The curve with label ``4" is the calculated braking torque caused by the nuclear expansion
of the non-degenerate star. The lower panel is a zoom-in on the upper two curves. 

It can be seen that if the magnetic braking law is relatively weak at short orbital periods
(Stepiens law) then the braking torque due to nuclear expansion becomes more important
than the magnetic stellar wind braking. It is therefore important to include this spin-orbit
effect in all detailed calculations of close binaries with a low-mass star.

\section{Discussion}
Whether or not the actual magnetic braking torque is weak or strong at short 
orbital periods is, in our opinion, not a settled question. Despite the emerging evidence
for a saturation of the magnetic braking torque at short orbital periods, we find that
there are problems with the formation of short orbital period (0.5--5 days) binary
millisecond pulsars with low-mass ($0.2\,M_{\odot}$) helium white dwarf companions
{\em if} the magnetic braking torque is weak. The main outline of the controversy is
the following (see our publication to be submitted shortly): 
There is emerging evidence (observationally and partly theoretically,
see Section 1) that the effects of a magnetic stellar wind saturates a short
orbital periods. On the other hand, a weak magnetic braking law 
(fx. Stepien 1995) results in a timescale problem
since the required formation time for a binary millisecond pulsar, with the above
mentioned observed properties, becomes larger than the Hubble-time ($\sim$ 13 Gyr).
The total formation time is the sum of the pre-RLO evolution time of the 
LMXB progenitor, plus the mass transfer time in the {X}-ray emitting LMXB phase.
The problem can not just be solved by increasing the donor star mass, since
the final mass of the white dwarf companion to the millisecond pulsar
has to be $\sim 0.20\,M_{\odot}$ or less.

\section{Conclusions}
We have demonstrated that the nuclear expansion of a (sub)giant star in a
close binary will cause the orbital period to decrease by $\sim$ 10$\,$\%
prior to RLO. The magnitude of this effect is similar what is expected
in the case of a weak magnetic braking law. Hence, we recommend all
detailed binary evolution calculations to include the effect of 
nuclear expansion. The strength of the actual magnetic braking is uncertain
and we encourage further investigations.

\acknowledgments
I would like to thank GertJan Savonije for many discussions and all
the participants at the conference (especially those who provoked me
to calculate this effect in detail).

\end{document}